\documentclass[12pt]{article}
\usepackage{amsmath}
\usepackage{graphicx}
\usepackage{subcaption}
\usepackage{enumerate}
\usepackage{natbib}
\usepackage{hyperref}
\usepackage{amssymb}
\usepackage{diagbox}
\usepackage{float}
\usepackage{url} 
\usepackage{setspace}
\usepackage{multirow}
\usepackage{makecell}
\usepackage{booktabs}

\usepackage{amsthm}
\newtheorem{prop}{Proposition}

\newcommand{\blind}{1}

\newcommand{\cm}[1]{\ignorespaces}

\def\bfd{\mathbf d}

\def\bfx{\mathbf x}

\def\bfz{\mathbf z}

\def\bfB{\mathbf B}

\def\bfI{\mathbf I}
\def\bfX{\mathbf X}

\def\bfalpha{\boldsymbol \alpha}
\def\bfbeta{\boldsymbol \beta}
\def\bftheta{\boldsymbol \theta}

\def\bfzero{\boldsymbol 0}

\def\bfTheta{{\ensuremath\boldsymbol{\Theta}}}

\def\diag{\mathrm{diag}}

\DeclareMathOperator*{\argmin}{arg\,min}

\newcommand{\mymatrix}[2]{\mathcal{M}_{#1\times #2}(\mathbb{R})}

\begin{document}

\def\spacingset#1{\renewcommand{\baselinestretch}%
{#1}\small\normalsize} \spacingset{1}

\newcommand{\mytitle}{\bf Bayesian Inference on Brain-Computer Interfaces via GLASS}

\if1\blind
{
  \title{\mytitle}
  \author{Bangyao Zhao\\
    Department of Biostatistics, University of Michigan\\
    and \\
    Jane E. Huggins \\
    Department of Physical Medicine and Rehabilitation and\\Department of Biomedical Engineering, University of Michigan\\
        and \\
    Jian Kang\thanks{To whom correspondence should be addressed: Jian Kang (jiankang@umich.edu)} \\
    Department of Biostatistics, University of Michigan}
  \maketitle
} \fi

\if0\blind
{
  \bigskip
  \bigskip
  \bigskip
  \begin{center}
 
{\LARGE \mytitle}
\end{center}
  \medskip
} \fi

\bigskip
\begin{abstract}
Brain-computer interfaces (BCIs), particularly the P300 BCI, facilitate direct communication between the brain and computers. The fundamental statistical problem in P300 BCIs lies in classifying target and non-target stimuli based on electroencephalogram (EEG) signals. However, the low signal-to-noise ratio (SNR) and complex spatial/temporal correlations of EEG signals present challenges in modeling and computation, especially for individuals with severe physical disabilities—BCI's primary users. 
To address these challenges, we introduce a novel \textbf{G}aussian \textbf{La}tent channel model with \textbf{S}par\textbf{s}e time-varying effects (GLASS) under a fully Bayesian framework. GLASS is built upon a constrained multinomial logistic regression particularly designed for the imbalanced target and non-target stimuli. The novel latent channel decomposition efficiently alleviates strong spatial correlations between EEG channels, while the soft-thresholded Gaussian process (STGP) prior ensures sparse and smooth time-varying effects. We demonstrate GLASS  substantially improves BCI's performance in participants with amyotrophic lateral sclerosis (ALS) and identifies important EEG channels (PO8, Oz, PO7, and Pz) in parietal and occipital regions that align with existing literature. For broader accessibility, we develop an efficient gradient-based variational inference (GBVI) algorithm for posterior computation and provide a user-friendly Python module available at \if1\blind{\url{https://github.com/BangyaoZhao/GLASS}}\fi\if0\blind{\url{https://github.com}}\fi. 
\end{abstract}

\noindent%
\vfill
\setcounter{page}{0}
\thispagestyle{empty}
\newpage
\spacingset{1.9} 
\section{Introduction}
\label{sec:intro}


A brain-computer interface (BCI) facilitates direct communication between a user and a computer using electroencephalogram (EEG) signals. BCIs have shown great potential in helping individuals with severe physical impairments to communicate with the world, notably for amyotrophic lateral sclerosis (ALS) patients. However, dysregulated EEG signals commonly found in ALS patients often compromise the efficacy of BCIs, and few methods exist to mitigate this issue. Our research specifically concentrates on improving the performance of the P300 BCI for ALS patients. 

\begin{figure}
    \centering
    \includegraphics[width=1\linewidth]{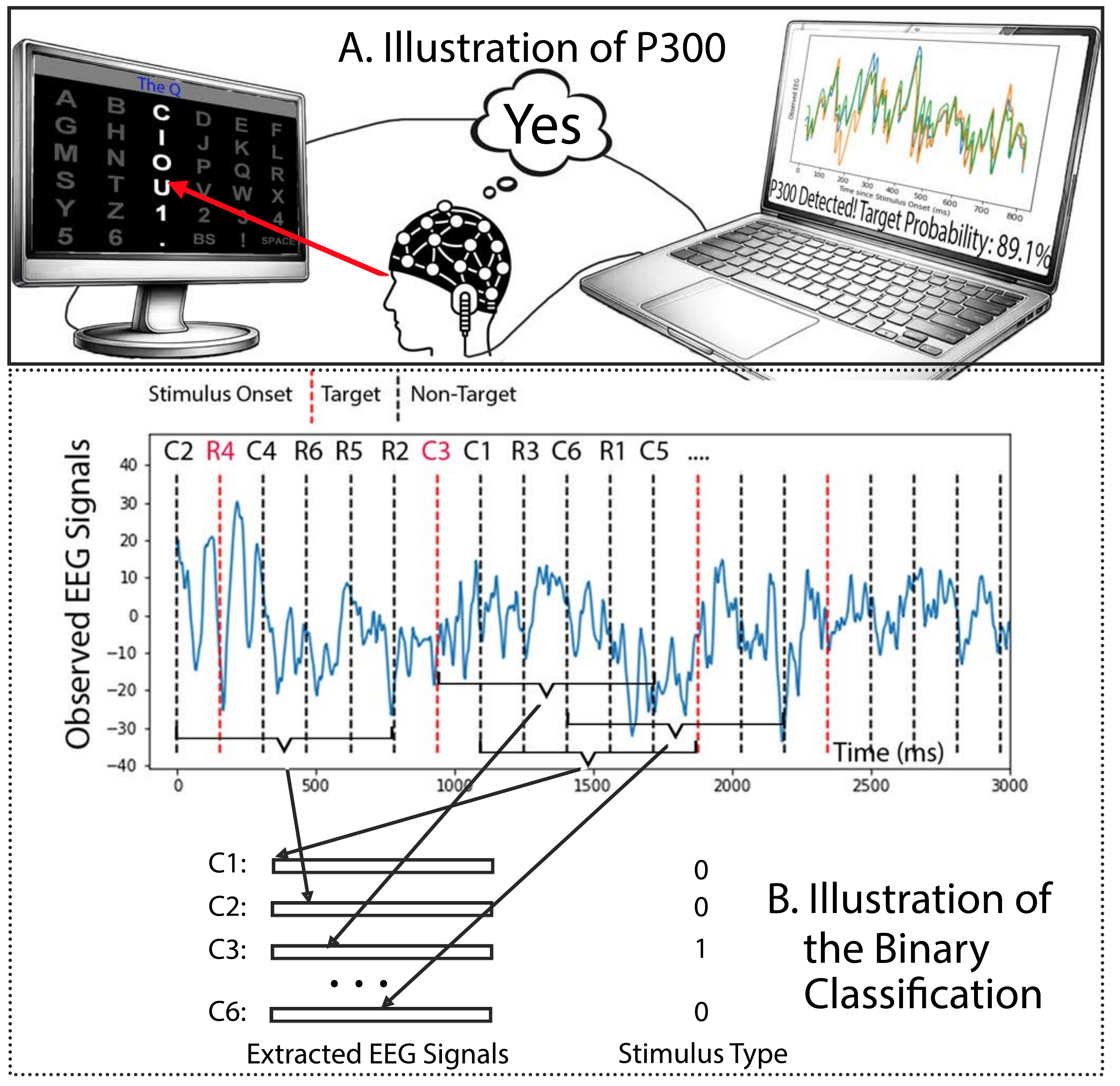}
    \caption{Panel A: Illustration of the P300 BCI. Displayed is a $6\times6$ virtual keyboard, where the rows and columns flash in a randomly permuted order.  In this example, the participant aims to type the character ``U". The current stimulus (flash) is on column 3 and is a target stimulus. Panel B: Illustrations of the binary classification framework. The labels ``C3 R4 C4 ..." (column 3, row 4, column 4, ...) at the top represent the order of stimulus presentation. 800 ms of EEG signals are extracted following each stimulus. The binary classification frameworks make independent predictions of stimulus types based on extracted EEG signals.}
    \label{fig:p300_speller_illustration}
\end{figure}

The P300 BCI, first proposed by \cite{farwell1988talking}, utilizes the event-related potential (ERP) for character typing. The name ``P300" refers to a distinct EEG component of the ERP in response to the presentation of a rare or meaningful stimulus. This component displays a positive peak in amplitude around 300 milliseconds post-stimulus and is thus called the P300 ERP. Panel A of Figure~\ref{fig:p300_speller_illustration} illustrates the P300 BCI. During the task, participants focus on a target character on a $6\times6$ grid keyboard. The system flashes the rows and columns randomly. Each flash is considered a stimulus, and the stimulus is called a target stimulus if it contains the target character. Each sequence has two target stimuli and ten non-target stimuli. Participants are asked to make mental responses to target stimuli, e.g., saying ``yes" mentally, with no physical responses needed. The computer monitors real-time EEG signals through multiple electrodes to determine whether the current stimulus is a target. The core computational neuroscience challenge in P300 BCIs is accurately classifying each stimulus as either a target or non-target. Correct binary classification allows for the precise identification of the target row and column, which ultimately identifies the target character at their intersection. In practice, a fixed length (typically 800 milliseconds) of EEG signals are extracted following each stimulus (see Panel B of Figure~\ref{fig:p300_speller_illustration}). Extracted EEG signals are then used to predict each stimulus type (target/non-target), which effectively formulates the task as a binary classification problem. 

There is a wide range of classification methods in P300 BCI applications, including the stepwise linear discriminant analysis (swLDA) \citep*{farwell1988talking,mowla2017enhancing, thompson2012classifier}, support vector machine (SVM) \citep{kaper2004bci}, ordinary least-squares (OLS) regression \citep{krusienski2006comparison}, logistic regression \citep{sakamoto2009supervised}, and the extreme gradient boosting (XGBoost) \citep*{sarraf2023study}. Since 2018, there has been a marked surge in neural network-based methodologies. Notable examples include a compact convolutional neural network named EEGNet \citep{lawhern2018eegnet}, a convolutional long short-term memory (ConvLSTM) model \citep{joshi2018single}, and a multi-task autoencoder-based model \citep{ditthapron2019universal}. \citet{ma2022bayesian} uniquely adopts a Bayesian generative framework to model EEG patterns using the Gaussian process (GP) prior, offering enhanced interpretability of brain signals. However, despite many of these methods being applied to ALS patients, they seldom address the unique challenges posed by EEG dysregulation in ALS patients. 

The classification of stimuli types poses several challenges. ALS Patients  often show reduced ERPs due to cognitive impairments, display abnormal EEG oscillatory activity, and struggle with sustained and selective attention. These factors contribute to lower signal-to-noise ratio (SNR) in ALS patients compared with healthy participants. Also, EEG signals are commonly sampled at frequencies more than 200 Hz and captured by multiple channels, leading to a high-dimensional feature space. For example, a typical EEG recording at 256 Hz during an 800-millisecond interval generates 205 time points, and with a 16-channel EEG cap, the feature space dimension increases to $205\times16=3280$. Moreover, EEG signals have strong spatial correlations due to the conductive properties of the skull and brain tissues, and they also exhibit high temporal correlations reflecting continuous brain activities. The presence of strong correlations may lead to overfitting problems. Furthermore, the study design includes only two target stimuli within a 12-stimulus sequence, resulting in an imbalanced dataset.

This article proposes a novel Bayesian approach for P300 BCI classification: the Gaussian latent channel model with sparse time-varying effects (GLASS). GLASS is built upon a constrained multinomial logistic regression model that fundamentally differs from existing binary classification methods. This framework naturally solves the dataset imbalance by directly identifying the target stimulus from the six row/column stimuli. With the goal of enhancing the SNR for ALS patients, we develop several novel prior specifications under a Bayesian framework. First, we construct a latent channel decomposition of the time-varying regression coefficients. This latent channel, constructed as a linear combination of original EEG channels, is designed to cancel out oscillations and noises between channels, moderate spatial correlations, and thereby improve the SNR. Second, we apply the soft-thresholded Gaussian Process (STGP) prior to impose sparsity on the time-varying coefficients. STGP supports continuously varying regression coefficients, enables feature selection, and leads to posterior consistency with high-dimensional data \citep*{kang2018scalar}. These properties are highly desired in EEG-BCI applications. For posterior computation, we utilize a gradient-based variational inference (GBVI) method to estimate the posterior distribution of model parameters and simulate the posterior predictive distribution of the target stimuli.

The remainder of the article is organized as follows. Section \ref{sec: method} develops the GLASS model and prior specifications, as well as the GBVI algorithm for posterior computation. Section~\ref{sec: application} presents the analysis of recorded BCI data using GLASS in comparison to existing methods. Section \ref{sec: simulation} performs extensive simulations to demonstrate the classification accuracy and robustness of GLASS. Finally, concluding remarks are given in Section \ref{sec: conc}.

\section{Method: GLASS}\label{sec: method}
\label{sec:meth}


We begin by introducing necessary notations. Let $c\ (c=1,\dots,C)$ index characters, and for each character suppose there are $S$ sequences (indexed by $s\ (s=1,\dots,S)$). Each sequence contains two half-sequences corresponding to row stimuli and column stimuli. We use $u\ (u=1,2)$ to denote the half-sequence type, with $u=1$ and $u=2$ representing row and column half-sequences, respectively. Stimuli in a half-sequence are indexed by the row/column number $j\ (j=1,\dots,6)$. For example, $(c, s, u, j)=(2,3,1,5)$ represents the stimulus on row five in the third sequence generated for typing the second character. To reduce notation complexity, we define $\mathcal{K} = \{(c, s, u)|\ c=1,\ldots,C;\ s=1,\ldots,S;\ u = 1,2\}$. We index elements in $\mathcal{K}$ using $i\ (i=1,\dots,N)$, so that every $\kappa_i\in\mathcal{K}$ uniquely determines a half-sequence. Here, $N=2CS$ is the total number of half sequences. 

Due to the nature of the P300 design, each half-sequence contains only one target stimulus and five non-target stimuli. Let $y_{ij}\in\{0, 1\}$ denote the stimulus type (target as 1 and non-target as 0) of stimuli $(i, j)$, i.e., the stimulus on row/column $j$ in half-sequence $i$. We have $\sum_{j=1}^6y_{ij}=1$ for every $i=1,\ldots,N$. Let $z_i\ (z_i\in \{1,\ldots, 6\})$ denote the row/column number of the target stimulus in half-sequence $i$, hence $z_i = j$ is equivalent to $y_{i j}=1$ and $y_{i j'} = 0$ for all $j' \neq j$.

After each stimulus, we collect EEG signals within a predetermined time window, typically 800 milliseconds, at a set of discrete time points $\{t_m|m=1,\ldots, M\}$, where we assume $0=t_1 <\ldots< t_{M}$. Let $x_{ijem}\in \mathbb{R}$ be the EEG signal from channel $e\ (e=1,\dots, E)$ at time $t_m$ following stimuli $(i, j)$. Let $\bfx_{ije}=[x_{ije1},\ldots,x_{ijem}]^\top\in\mathbb{R}^M$ and $\bfX_{ij}=[\bfx_{ij1}^\top,\ldots,\bfx_{ijE}^\top]^\top\in \mathcal{M}_{E\times M}(\mathbb{R})$. Here, $\mathcal{M}_{E\times M}(\mathbb{R})$ denotes the set of all $E\times M$ real matrices. Intuitively, $\bfX_{ij}$ is the matrix of EEG signals following stimuli $(i, j)$, with each row corresponding to a channel and each column corresponding to a time point. 


\subsection{Constrained Multinomial Logistic Regression Model}

Let $\bfX_{i} = [\bfX_{i 1}^\top,\ldots,\bfX_{i 6}^\top]^\top\in\mymatrix{6E}{M}$. The aim is to specify the distribution of $z_i$ given $\bfX_i$ for each half-sequence $i$.
\begin{align}
\mathrm{Pr}(z_i = j|\bfX_i) &= \frac{\exp(\eta_{ij})}{\sum_{j'=1}^6 \exp(\eta_{ij'})}\quad \mbox{with}\quad \eta_{ij}=\langle\bfX_{ij},\bfB\rangle\quad \mbox{for}\quad j=1,\ldots,6\label{eq: linear}.
\end{align}
Here, $\langle\cdot,\cdot\rangle$ represents the inner product of two matrices of the same size. Parameter $\bfB \in \mymatrix{E}{M}$ is the matrix of regression coefficients.  In \eqref{eq: linear}, the probability of $z_i = j$ is linked to $\eta_{ij}$ by the softmax function, and $\eta_{ij}$ is interpreted as the un-normalized log-odds of stimulus $j$ being the target in half-sequence $i$. The $(e, m)$-th element of $\bfB$, denoted as $\beta_{em}$, is the time-varying effect of channel $e$ at time $t_m$. 


The formulation in \eqref{eq: linear} is a constrained version of the general multinomial logistic regression, which is expressed as $\eta_{ij}=\sum_{j'=1}^6\langle\bfX_{ij'},\bfB_{jj'}\rangle$. Here, $\bfB_{jj'}\in \mymatrix{E}{M}$ is the matrix of regression coefficients for predicting the type of stimulus $j$ specific to EEG signals following stimulus $j'$. We put two constraints on the general form, i) $\bfB_{jj'}=\bfzero$ if $j\neq j'$, i.e., EEG signals following the stimulus $j'$ do not predict the type of stimulus $j$, and ii) $\bfB_{11}=\ldots=\bfB_{66}:=\bfB$, implying that the model does not differentiate which row/column the stimulus is on. 

We derive the sufficient and necessary identifiability condition for the proposed constrained multinomial logistic regression. Let $\Vec(\cdot)$ be the vectorization function, which transforms a matrix into a column vector by stacking its columns one after the other. Let $\bfX^*_i=[\Vec(\bfX_{i 1}-\bfX_{i 6}),\ldots,\Vec(\bfX_{i 5}-\bfX_{i 6})]^\top\in\mymatrix{5}{EM}$ and $\bfX^*=[\bfX^{*\top}_1,\ldots,\bfX^{*\top}_N]^\top\in\mymatrix{5N}{EM}$.

\begin{prop}
The model defined by~\eqref{eq: linear} is identifiable if and only if $\bfX^*$ is of full column rank.
\end{prop}
\begin{proof} For each $i$, $[\eta_{i 1}-\eta_{i 6},\ldots,\eta_{i 5}-\eta_{i 6}]^\top\in \mathbb{R}^5$ uniquely determines the distribution of $z_i$. Therefore, the identifiability condition is equivalent to the linear equation system of
$\langle\bfX_{ij}-\bfX_{i6},\bfB\rangle=\eta_{ij}-\eta_{i6}$,
for $i=1,\ldots,N$ and $j=1,\ldots,5$, having a unique solution. 

Suppose $\bfbeta = \Vec (\bfB)\in\mathbb{R}^{EM}$, $\bfd_i=[\eta_{i1}-\eta_{i6},\ldots,\eta_{i5}-\eta_{i6}]^\top\in\mathbb{R}^5$, and $\bfd = [\bfd_1^\top,\ldots,\bfd_N^\top]^\top\in\mathbb{R}^{5N}$. The linear equation system can be represented in matrix form: $\mathbf{X^*}\bfbeta=\bfd$. 
Therefore, the model is identifiable if and only if the design matrix $\mathbf{X^*}$ has a full column rank. 
\end{proof}

\subsection{Prior Specifications}

\subsubsection{Latent Channel Decomposition of Time-varying Effects}

When there are no constraints on the matrix of coefficients $\bfB$, Proposition 1 implies a necessary condition for model identifiability, which is the number of rows in $\mathbf{X^*}$ being no less than the number of columns, i.e., $5N\geq EM$. Here, $5N$ is $5/6$ of the total number of stimuli, and $EM$ is the number of EEG features following each stimulus. In practice, $EM$ can be larger than $5N$ due to the high frequency of EEG data, which can make the model unidentifiable. Moreover, the strong spatial correlations among EEG channels may lead to computational issues. To address these issues, we propose a latent channel decomposition of the time-varying effects, which is
\begin{equation}\bfbeta_e=\delta_{e}\alpha_{e}\Tilde{\bfbeta}\quad \mbox{for}\quad e=1,\ldots, E\label{eq: group},
\end{equation}
where $\bfbeta_e=[\beta_{e1},\ldots,\beta_{eM}]^\top\in\mathbb{R}^M$ is the $e$-th row of the coefficient matrix $\bfB$ as a column vector, $\Tilde{\bfbeta}=[\Tilde{\beta}_1, \ldots,\Tilde{\beta}_M]^\top\in\mathbb{R}^M$ is the vector of latent channel time-varying effects, $\delta_e\in \{0, 1\}$ is the selection indicator of channel $e$, and $\alpha_{e} \in [-1,1]$ is the contribution weight of the latent channel time-varying effects to channel $e$. We assume that the contributing weights are normalized with the $L_2$-norm, i.e., $\sum_{e=1}^E \alpha^2_{e} = 1$. This decomposition is equivalent to approximating the matrix of time-varying effects $\bfB$ using the rank-one matrix $\Delta\bfalpha \Tilde{\bfbeta}^\top$, where $\bfalpha = [\alpha_{1},\ldots,\alpha_{E}]^\top$ and $\Delta = \diag(\delta_1,\ldots, \delta_E)$.  Thus, with this decomposition, the number of parameters is reduced in the model. 

Equations \eqref{eq: linear} and \eqref{eq: group} can be equivalently expressed as two steps, i) summarizing $E$ channels of EEG signals into a single latent channel $\Tilde{\bfx}_{ij}=\sum_{e=1}^E\delta_{e}\alpha_{e}\bfx_{ije}$ and ii) re-interpreting $\Tilde{\bfbeta}$ as new regression coefficients, i.e., $\eta_{ij}=\Tilde{\bfx}_{ij}^\top\Tilde{\bfbeta}$. Therefore, the selection indicator $\delta_e$ can be considered as an indicator of whether channel $e$ contributes to the latent channel, while $\alpha_e$ can be considered as the contribution weight of EEG signals from channel $e$ to the latent channel. This latent channel effectively preserves most information from original channels while mitigating channel-wise correlations in EEG signals.


\subsubsection{Soft-thresholded GP Prior with Global Shrinkage}
To further improve the SNR, we propose a soft-thresholded GP prior with global shrinkage for latent channel time-varying effects. Let $S_\tau(\cdot)$ be the soft thresholding function with a threshold parameter $\tau > 0$, i.e., $S_\tau(x)= \mathrm{sign}(x)(|x|-\tau)I(|x|>\tau)$. Let $\mathcal{N}(\mu,\sigma^2)$ denote the normal distribution with mean $\mu$ and variance $\sigma^2$, and $C^+(A)$ denote a half-Cauchy distribution with the scale parameter $A$. For $m = 1,\ldots, M$, we assume
\begin{equation}
\Tilde{\beta}_m=S_\tau(\beta^*_m), \quad \beta^*_m|\beta^*_{m-1} \sim \mathcal{N}(\beta^*_{m-1} , \sigma^2), \quad \sigma \sim C^+(A)
\quad \mbox{for}\quad m=1,\ldots, M,\label{eq: soft thresholding}
\end{equation}
where $\beta^*_m$ is the raw time-varying effect before soft-thresholding and $\beta^*_0$ is fixed at zero. The variance parameter $\sigma^2$ imposes a global shrinkage on the difference in the raw time-varying effect between time points $t_m$ and $t_{m-1}$ towards zero, which ensures the time continuity of $\Tilde{\bfbeta}$. The tuning parameter $A$ controls the level of shrinkage imposed on the difference of time-varying effects. 

\subsubsection{Latent Channel Projected Normal Prior}

We present prior specifications for the selection indicators and contribution weights. Let $\mathrm{Bern}(p)$ denote a Bernoulli random variable with probability $p$.  Let $\bfalpha=[\alpha_{1},\ldots,\alpha_{E}]^\top$. Then we have
\begin{align}
    \delta_e\sim \mathrm{Bern}\left(0.5\right),\quad
    \bfalpha= \bfalpha^*/\|\bfalpha^*\|_2,\quad \bfalpha^*\sim \mathcal{N}(0,\bfI_E),
    \label{eq: projection}
\end{align}
where the prior probability of selecting channel $e$ is 0.5, and $\bfalpha$ follows a projected normal distribution~\citep{wang2013directional} with the $L_2$-norm being one. This ensures the scale of the time-varying effects at the latent channels is comparable to that at the original EEG channels and thus ensures the identifiability of GLASS. 

\begin{figure}
    \centering
    \includegraphics[width=0.8\linewidth]{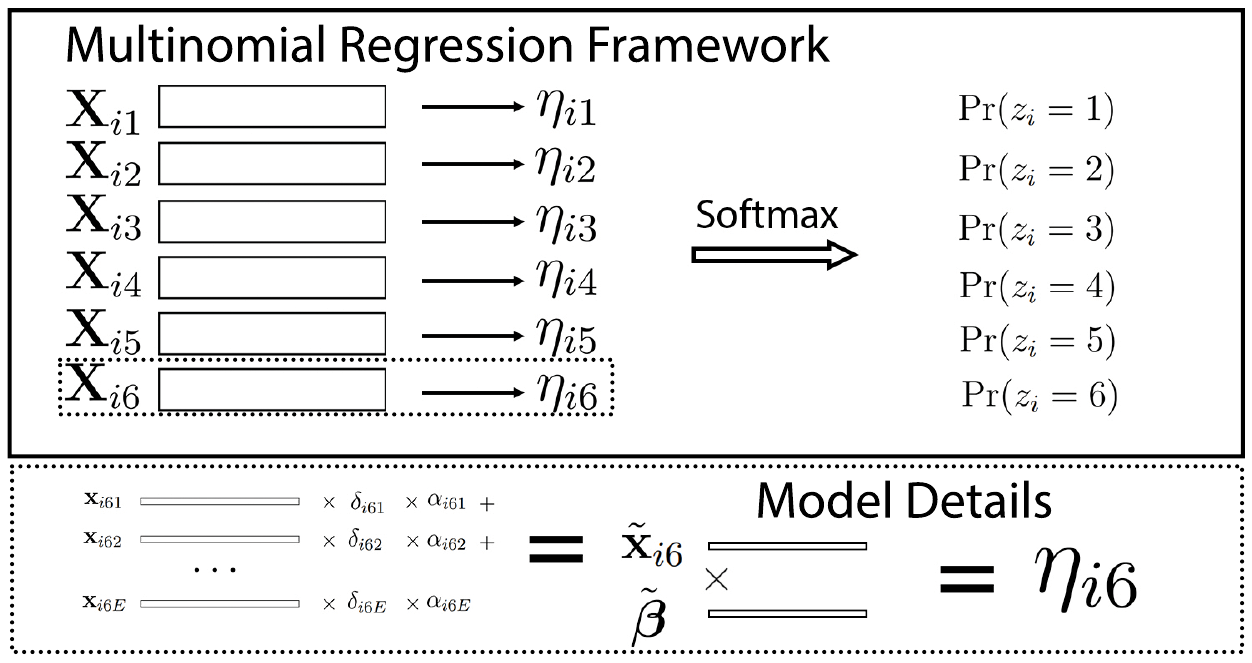}
    \caption{Illustration of GLASS. The first panel illustrates the multinomial regression framework. For each half-sequence, the model calculates unnormalized log odds for each row/column based on extracted EEG signals. These log odds are normalized into probabilities of being the target using the softmax function. The second panel illustrates how unnormalized log odds are calculated for each stimulus, using the stimulus on row/column 6 as an example. First, based on selection indicators and contribution weights, GLASS summarizes multiple EEG channels into one latent channel. Then, signals in the latent channel are multiplied with the latent channel time-varying effects, which gives the unnormalized log odds.}
    \label{fig: model summary}
\end{figure}

Figure \ref{fig: model summary} summarizes the model specifications of GLASS. The constrained multinomial logistic regression modeling framework is presented in \eqref{eq: linear}; Equations \eqref{eq: group}, \eqref{eq: soft thresholding}, and \eqref{eq: projection} give the prior specifications. Let $\bftheta=[\beta^*_1,\ldots,\beta^*_M,\sigma, \delta_1,\ldots,\delta_E,\alpha^*_1,\ldots,\alpha^*_E]^\top$ be the collection of all the model parameters and  $\bfTheta$ represents the corresponding parameter space. Let $\bfz=[z_1, \ldots,z_N]^\top$ be the vector of observed outcomes and $\pi(\bftheta)$ represent the prior density function. Let $\pi(\bfz|\bftheta,\bfX_1,\ldots,\bfX_N)$ be the likelihood function and $\pi(\bfz|\bfX_1,\ldots,\bfX_N)=\int_{\bfTheta}\pi(\bfz,\bftheta|\bfX_1,\ldots,\bfX_N)d\bftheta$ be the marginal likelihood, where $\pi(\bfz,\bftheta|\bfX_1,\ldots,\bfX_N)$ denotes the joint density function and $\pi(\bftheta|\bfz,\bfX_1,\ldots,\bfX_N)$ is the posterior density. 

\subsection{Posterior Computation}\label{sec: Posterior Computation}

We perform gradient-based variational inference (GBVI) for posterior computation. GBVI combines the power of stochastic optimization and variational inference. GBVI can generally provide accurate point estimations for model parameters and is faster than sampling methods, including Gibbs samplers. The variational inference approximates the true posterior $\pi(\bftheta|\bfz, \bfX_1,\ldots,\bfX_N)$ by seeking a ``surrogate" distribution (denoted by $q^*$) that minimizes the KL-divergence with the true posterior within a distribution family $\mathcal{Q}$ \citep{blei2017variational}, i.e., $q^*=\argmin_{q\in \mathcal{Q}}\mathrm{KL}(q||\pi)$, 
where $\mathrm{KL}(q||\pi)=\mathbb{E}_q[\log q(\bftheta)] - \mathbb{E}_q[\log \pi(\bftheta|\bfz,\bfX_1,\ldots,\bfX_N)]$. The computation of KL-divergence is difficult as it involves the intractable normalizing constant. Minimizing the KL-divergence is equivalent to maximizing the evidence lower bound (ELBO), which is defined as 
$\mathrm{ELBO}(q)=\mathbb{E}_q[\log \pi(\bftheta, \bfz|\bfX_1,$ $\;\ldots,\bfX_N)]-\mathbb{E}_q[\log q(\bftheta)]$. 
Suppose the distribution family $\mathcal{Q}$ is parameterized by $\xi\in U$, where $U$ is an open subset of Euclidean space, then $\mathrm{ELBO}(q)$ can be regarded as a function of $\xi$. To perform gradient-based algorithms, various gradient estimators have been proposed to approximate $\nabla_\xi \mathrm{ELBO}(q)$, including i) the score estimator (also known as the log-derivative gradient estimator) \citep{williams1992simple}, ii) an extension of i) with reduced variance \citep{mnih2016variational}, iii) reparameterization gradients as introduced by \cite{kingma2013auto}, and iv) doubly-reparameterized estimator presented by \cite{tucker2018doubly}. In our application study, we adopt iii) reparameterization gradients and the optimizer Adam due to its scalability to large data \citep{kingma2014adam}. 



\subsubsection{Surrogate Distributions}
For GLASS, we construct the surrogate distribution as the product of independent distributions. GLASS is fully parameterized by the following parameters, and the distribution family for each parameter is specified as the following.  
\begin{itemize}
    \item $\beta^*_m$'s: normal distributions, parameterized by means and variances. 
    \item $\sigma$: log-normal distributions, parameterized by the normal mean and variance.
    \item $\delta_e$'s: Bernoulli distributions parameterized by log-odds. 
    \item $\alpha^*_{e}$'s: normal distributions, parameterized by means and variances. 
\end{itemize}
We use the softplus function, i.e., $f(x)=\log[1 + \exp(x)]$, to map variances to the entire real line to ensure numerical stability of the computation. 


\subsection{Prediction}\label{sec: prediction}
For a new half-sequence $i'$, we derive the posterior predictive distribution of the target row/column number $z_{i'}$ given the EEG matrix $\bfX_{i'}$. Based on \eqref{eq: linear}, the distribution of $z_{i'}$ given a fixed model parameter $\bftheta$ can be expressed as
\begin{align}
\mathrm{Pr}(z_{i'} = j|\bfX_{i'}) &= \frac{\exp(\eta_{i'j})}{\sum_{j'=1}^6 \exp(\eta_{i'j'})}\quad \mbox{with}\quad \eta_{i'j}=\langle\bfX_{i'j},\bfB\rangle\quad \mbox{for}\quad j=1,\ldots,6.
\end{align}
We adopt an importance sampling approach to approximate the posterior predictive distribution of $z_{i'}$ given $G$, say $G=2000$, i.i.d. samples from $q$, denoted as $\{\bftheta_g|g=1,\ldots, G\}$.
\begin{align*}
&\mathrm{Pr}(z_{i'}=j|\bfz, \bfX_1,\ldots,\bfX_N,\bfX_{i'})\\
=&\int_{\bfTheta} \mathrm{Pr}(z_{i'}=j|\bftheta,\bfz, \bfX_1,\ldots,\bfX_N,\bfX_{i'})\pi(\bftheta|\bfz, \bfX_1,\ldots,\bfX_N,\bfX_{i'})d\bftheta\\
=&\int_{\bfTheta} \mathrm{Pr}(z_{i'}=j|\bftheta,\bfX_{i'})\pi(\bftheta|\bfz, \bfX_1,\ldots,\bfX_N)d\bftheta\label{eq: predictive dist}\\
=& \int_{\bfTheta} \mathrm{Pr}(z_{i'}=j|\bftheta,\bfX_{i'})\frac{\pi(\bftheta,\bfz| \bfX_1,\ldots,\bfX_N)}{q(\bftheta)\pi(\bfz| \bfX_1,\ldots,\bfX_N)}q(\bftheta)d\bftheta\\
\approx& \frac{1}{\pi(\bfz| \bfX_1,\ldots,\bfX_N)M}\sum_{g=1}^G\mathrm{Pr}(z_{i'}=j|\bftheta_g, \bfX_{i'})\frac{\pi(\bftheta_g,\bfz| \bfX_1,\ldots,\bfX_N)}{q(\bftheta_g)}
\end{align*}
Here, $\mathrm{Pr}(z_{i'}=j|\bftheta,\bfz, \bfX_1,\ldots,\bfX_N,\bfX_{i'})=\mathrm{Pr}(z_{i'}=j|\bftheta,\bfX_{i'})$ and $\pi(\bftheta|\bfz, \bfX_1,\ldots,\bfX_N,\bfX_{i'})=\pi(\bftheta|\bfz, \bfX_1,\ldots,\bfX_N)$ are two conditional independence assumptions. Notice that $\pi(\bfz| \bfX_1,\ldots,\bfX_N)$ in the last equation is a normalizing constant. In computation, we ignore $\pi(\bfz| \bfX_1,\ldots,\bfX_N)$ and re-normalize the probabilities into one afterward. 

In practice, multiple (say $S'$) sequences are generated for one character. The $S'$ half-sequences (either for rows or columns) $i'_1,\ldots,i'_{S'}$ should have the same target row/column number, i.e., $z_{i'_1}=\ldots=z_{i'_{S'}}:=z$. To combine information from multiple half-sequences, we predict the target row/column number as the one that maximizes the following conditional probability. 
\begin{align*}
&\mathrm{Pr}(z=j|\bfz, \bfX_1,\ldots,\bfX_N, \bfX_{i'_1},\ldots,\bfX_{i'_{S'}},z_{i'_1}=\ldots=z_{i'_{S'}}:=z)\\
&\propto\prod_{s'=1}^{S'} \mathrm{Pr}(z_{i_{s'}'}=j|\bfz, \bfX_1,\ldots,\bfX_N,\bfX_{i'_{s'}})
\end{align*}

\section{Application}\label{sec: application}

We apply GLASS to multiple BCI datasets collected from a 68-year-old female ALS participant (Participant A). We mainly demonstrate two aspects of GLASS: (i) improved BCI performance with fewer training data and (ii) offering scientific insights into key brain regions and essential ERP response time windows. Several state-of-the-art methods are compared with GLASS, including swLDA, logistic regression, OLS, EEGNet, SVM, SMGP, and XGBoost.

\subsection{Data Aquisition}

In this research, Participant A is equipped with an EEG cap with 16 channels. This participant is seated 0.8 meters away from a 17-inch monitor, which displays a virtual keyboard arranged in a $6\times6$ grid. The placement of the 16 scalp electrodes adheres to the international 10–20 system, and the channel names are F3, Fz, F4, T7, C3, Cz, C4, T8, CP3, CP4, P3, Pz, P4, PO7, Oz, and PO8. The device records brain activity at a sampling rate of 256 Hz. During the experiment, each stimulus is highlighted for a duration of 31.25 milliseconds, followed by a 125-millisecond interval before the next stimulus flashes. There is a 3.5-second pause between the last stimulus for one character and the first stimulus for the subsequent character. 

The BCI task involves typing ten sentences. The initial sentence ``THE QUICK BROWN FOX" (comprising 19 characters including spaces) is used for training the model. For each character, 15 sequences are generated, and each sequence consists of 12 stimuli. After model training, the participant utilizes the model to complete nine free-typing tasks. Each free-typing task involves typing a sentence varying in length from 20 to 25 characters. During these free-typing tasks, each character is presented for six sequences. The data collected during the free-typing sessions is used as the testing dataset.

\subsection{Data Processing and Model Fitting}

We extract 800 milliseconds of EEG signal following each stimulus, corresponding to $M=205$ time points with a sampling rate of 256 Hz. For traditional binary classification methods, this translates to a feature space of dimension $205\times16$ (channels) $=3280$ with 19 (characters) $\times$ 15 (sequences) $\times$ 12 (rows and columns) = 3420 training samples, leading to overfitting issues. Following established practices, we preprocess the EEG data for swLDA, logistic regression, OLS, SVM, XGBoost, and SMGP using a Butterworth bandpass filter with a frequency range of 0.5Hz to 15Hz. Then, we down-sample the EEG data to 32Hz (25 time points in 800 milliseconds). However, for GLASS and EEGNet, we retain the original sampling rate and do not apply bandpass filtering, as these methods are designed to be robust against high-dimensional data and low SNR. 

For GLASS, we configure Adam with 2000 iterations and a step size of 0.05. For gradient approximation in each iteration, the number of Monte Carlo samples is set to $10$. To determine the soft-thresholding parameter $\tau$, we fit a baseline model with $\tau=0$. Following this, we compute the median absolute value of the estimated $\Tilde{\bfbeta}$ in the baseline model. The final value of $\tau$ for GLASS is then set to be half of this median absolute value. This approach allows us to determine the soft-thresholding parameter in a data-driven way.

For swLDA, we set the number of iterations to 64. The p-values for feature inclusion and exclusion are set at 0.1 and 0.15, respectively. We employ the implementation of OLS in the Python module statsmodels \citep{seabold2010statsmodels} and the implementation of logistic regression and SVM in the Python module scikit-learn \citep{pedregosa2011scikit}. EEGNet configuration aligns with the guidance from their code on Github: we employ a 32 batch size and set the Adam learning rate to 0.001. The training process takes 500 epochs, using the binary cross-entropy loss as the loss function. In SMGP, we use a $\gamma$-exponential kernel with a scaling parameter of 0.5 and a shape parameter of 1.8. The MCMC settings follow original specifications with 2000 iterations and 1000 burn-ins across three independent chains. In XGBoost, we configure a maximum tree depth of 5, a learning rate of 0.03, a feature subsampling rate of 80\%, and minimum loss reduction required to make a further partition 10. The XGBoost model was trained over 1000 boosting rounds.

All models are trained using a six-core 12-thread AMD Ryzen 5 3600X 3.80 GHz 16 GB RAM x64 computer without GPU acceleration. The training time of GLASS, EEGNet, and XGBoost is 40.0s, 41.8s, and 34.9s, respectively. The training time of swLDA, logistic regression, OLS, and SVM is 3.4s, 0.3s, 0.5s, and 4.3s, respectively. The model training time of SMGP is 1.7 hours when running three chains in parallel. 

\subsection{Bayesian Inference via GLASS}

\subsubsection{Identifying the Latent Channel}

\begin{figure}[ht]
    \centering
    \includegraphics[width=0.8\linewidth]{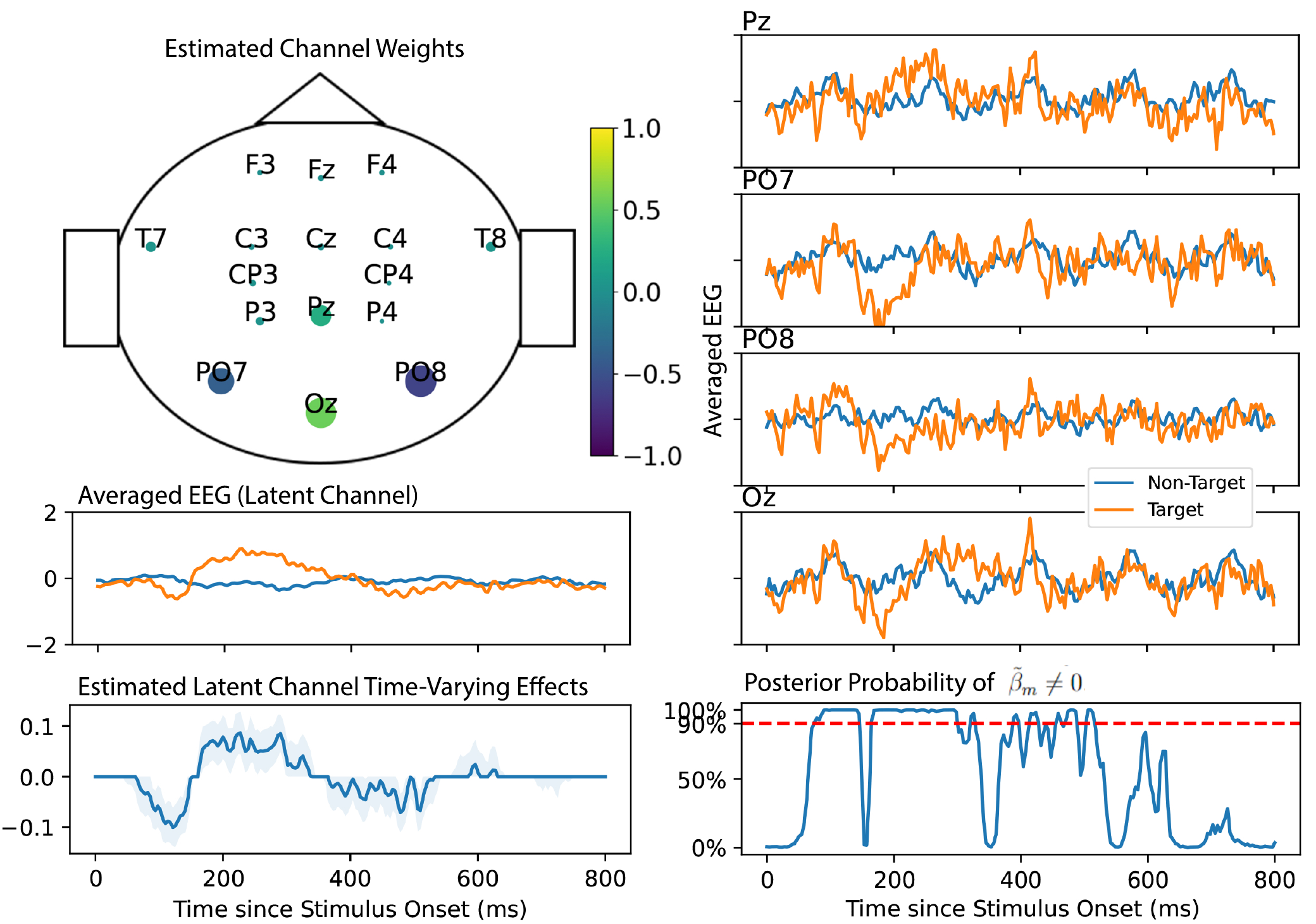}
    \caption{Upper left panel: Estimated channel weights (a top-down view); Middle left panel: Epoch-averaged 800ms EEG responses in the latent channel; Upper right panel: Epoch-averaged EEG responses from channels Pz, PO7, PO8, and Oz. Lower left panel: Estimated time-varying effects with 95\% credit intervals. Lower right panel: Posterior probability of $\Tilde{\beta}_m\neq0$.}
    \label{fig: weights}
\end{figure}

We calculate estimated channel weights as posterior medians. Figure (\ref{fig: weights}) shows estimated channel weights over electrode locations. Visually, channels Pz, PO7, PO8, and Oz exhibit prominent weights. We use the posterior mean selection probability to quantify the importance of channels. A channel is considered important if its posterior selection probability exceeds 90\%. Channels Pz, PO7, PO8, and Oz are identified as important channels, which align with the literature as they are located in the parietal and occipital regions. The two regions are important brain regions for visual processing and attention control \citep{takano2014coherent}. 

To highlight the enhanced SNR in the latent channel, we compute the averaged 800-ms EEG signals for target and non-target stimuli, respectively. Figure \ref{fig: weights} shows averaged EEG signals from PO7, PO8, Oz, Pz, and the latent channel. Visually, the latent channel has smoother EEG signals and a clearer separation of target/non-target responses compared to the original channels.

\subsubsection{Capturing ERP Response Windows}

We estimate $\Tilde{\bfbeta}$ using posterior medians and calculate 95\% credible intervals. For each time point $t_m$, we calculate the posterior probability of $\Tilde{\beta}_m\neq0$ to evaluate whether the ERP response is significant at $t_m$. Figure (\ref{fig: weights}) shows the estimated latent channel time-varying effects with 95\% credible intervals and the posterior probabilities of $\Tilde{\beta}_m\neq0$. The time interval near 100 ms is indicative of visual-evoked ERP responses, which typically require shorter response times. The subsequent time interval around 200-300 ms corresponds to the commonly known ``P300" ERP response. Interestingly, our method identifies a less pronounced response window near 400-500 ms, which has been rarely discussed in the literature, suggesting a new aspect for further research.

\subsection{BCI Performance Comparison} 

We assess the prediction accuracy of all methods using the testing dataset. For binary classification methods, we average the classification scores for each row and column across multiple sequences. The target character is then determined at the intersection of the row and column that yield the highest average scores. Section \ref{sec: prediction} has introduced how GLASS combines information from multiple sequences for target character identification. SMGP directly calculates the posterior probability of each key on the keyboard being the target character. Consequently, the character associated with the highest posterior probability is identified as the target character.

While prediction accuracy is important for BCIs, it is not the only metric of BCI performance. A more comprehensive metric, known as the BCI utility, is frequently used in many BCI applications \citep*{thompson2014plug}. This throughput metric quantifies the expected number of information bits correctly transferred through the system per unit time. We adopt the definition of BCI utility as outlined in \cite*{dal2009utility}. Given that different methods may achieve optimal BCI utility at different numbers of sequences, we calculate this metric as its maximum over the six sequences for a fair comparison. To evaluate the model performance with limited training data, we introduce the ``Less Training" scenario, where only the first three sequences of each character are used for model training, leading to an 80\% reduction in the size of the training dataset. This is in contrast to the ``Standard Training" scenario, which utilizes the entire training dataset to train the model. Table \ref{tab: accuracy} presents the means and standard deviations of both prediction accuracy and BCI utility across nine testing sentences, alongside the number of sequences required to attain an 80\% mean accuracy. Furthermore, we evaluate the performance of GLASS using the four selected channels alone (denoted as GLASSel). 

\begin{table}[h]
    \centering
    \caption{Means and standard deviations of prediction accuracy (\%) and BCI utility (bits/s) across nine testing sentences. The $n_{seq}^{80\%}$ represents the number of sequences needed for 80\% mean accuracy. ``-" denotes cases where a classifier never reached 80\% mean accuracy. The ``Standard Training" setting uses all training data available to train each model. We use the ``Less Training" setting to evaluate model performance with only 20\% of the training data. GLASSel represents GLASS using only the four selected channels PO7, PO8, Oz, and Pz.}
    \label{tab: accuracy}
    \begin{tabular}{c|c|c|c|c|c|c}
    \toprule
    &\multicolumn{3}{c|}{Standard Training}&\multicolumn{3}{c}{Less Training}\\\midrule
    Classifier&Accuracy&Utility&$n_{seq}^{80\%}$&Accuracy&Utility&$n_{seq}^{80\%}$\\\midrule
    GLASS&93.2\% (4.7\%)&21.6 (3.4)&3&93.1\% (6.3\%)&21.1 (5.6)&3\\\midrule
    \textbf{GLASSel}&\textbf{94.4\% (3.3\%)}&\textbf{22.0 (3.7)}&\textbf{2}&\textbf{93.5\% (3.7\%)}&\textbf{21.7 (4.1)}&\textbf{3}\\\midrule
    swLDA&89.9\% (6.8\%)&18.1 (4.5)&4&80.4\% (17.0\%)&11.9 (6.2)&6\\\midrule
    Logistic&85.3\% (7.4\%)&17.6 (4.9)&5&32.1\% (20.7\%)&0.0 (0.0)&-\\\midrule
    OLS&88.1\% (8.9\%)&15.7 (4.6)&4&58.5\% (18.9\%)&4.1 (7.5)&-\\\midrule
    EEGNet&89.7\% (5.8\%)&14.9 (4.4)&5&23.4\% (11.0\%)&0.0 (0.0)&-\\\midrule
    SVM&77.6\% (11.8\%)&11.3 (4.6)&-&61.2\% (9.6\%)&3.3 (1.9)&-\\\midrule
    SMGP&71.1\% (7.9\%)&6.5 (2.1)&-&68.8\% (9.2\%)&6.3 (2.5)&-\\\midrule
    XGBoost&62.7\% (13.5\%)&4.0 (4.2)&-&50.1\% (15.0\%)&1.8 (2.7)&-\\\bottomrule
    \end{tabular}
\end{table}

The results indicate that GLASS enhances BCI performance in both scenarios. This improvement is particularly marked when the amount of training data is limited. Remarkably, the performance of GLASS is slightly improved with only four selected channels. This underscores the value of GLASS in reducing the number of EEG electrodes needed, potentially leading to more cost-effective BCI systems.

\subsection{Sensitivity Analysis}

We conduct a sensitivity analysis to evaluate the impact of different choices of the soft-thresholding parameter $\tau$ on the performance of GLASS. In the initial study, we set $\tau$ as the shrinkage ratio of 0.5 times the median absolute value of the estimated $\Tilde{\bfbeta}$ in the baseline model with $\tau=0$. We explore a range of shrinkage ratios, including 0, 0.5, 1, and 2. The sensitivity analysis results under both ``Standard Training" and ``Less Training" scenarios are presented in Table \ref{tab: sensitivity}. The prediction accuracy and BCI utility exhibit only minor differences in response to different shrinkage ratios. This suggests that GLASS is not sensitive to different soft-thresholding parameters.

\begin{table}[h]
    \centering
     \caption{ Sensitivity analysis of GLASS's performance across various shrinkage ratios. In our application analysis, we set the shrinkage ratio at 0.5. We have two scenarios in the application study: ``Standard Training" and ``Less Training" scenarios. The ``Standard Training" scenario utilizes the full training dataset for training GLASS, whereas the ``Less Training" scenario employs only 20\% of the training data 
    }
    \begin{tabular}{c|c|c|c|c}\midrule
    &\multicolumn{2}{c|}{Standard Training}&\multicolumn{2}{c}{Less Training}\\\midrule
        Shrinkage Ratio & Accuracy&Utility & Accuracy&Utility \\\midrule
        0 & 92.5 (4.9)& 20.3 (4.2)& 93.5 (4.8)& 20.9 (4.1)\\\midrule
        0.5& 93.2 (4.7)&21.6 (3.4)&94.4 (3.3)& 22.0 (3.7)\\\midrule
        1& 92.8 (5.0)& 21.8 (4.1)& 92.8 (5.0)& 21.8 (4.1)\\\midrule
        2& 89.5 (7.3)& 17.9 (5.5)&90.1 (4.2)&19.1 (4.8)\\\midrule
    \end{tabular}
   
    \label{tab: sensitivity}
\end{table}

\subsection{Multi-Participant Analysis}

\begin{figure}[h]
    \centering
    \includegraphics[width=1\textwidth]{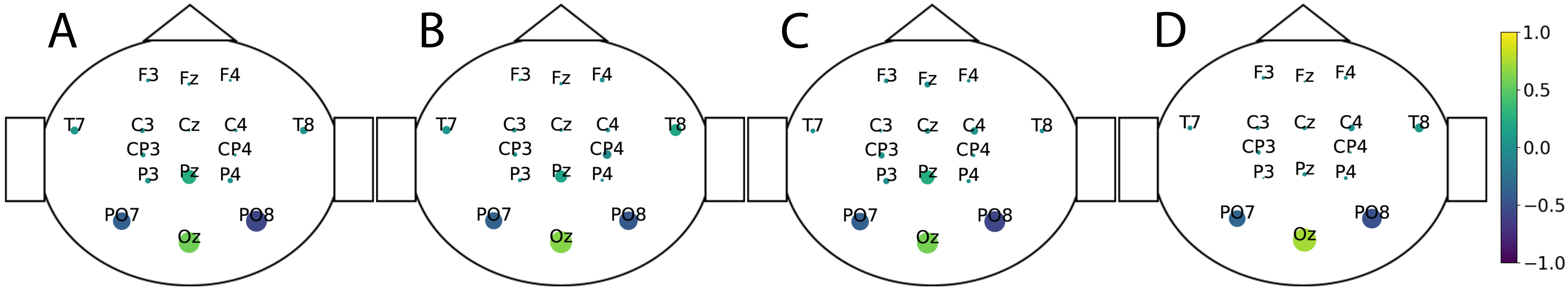}
    \caption{Estimated contribution weights of EEG channels, mapped over the electrode locations on the EEG cap, for Participants A, B, C, and D, respectively. }
    \label{fig: participants weights}
\end{figure}

Participant A, a 68-year-old female with ALS, is paired with Participant B, a 66-year-old male ALS patient, for an age-matched comparison. Additionally, two age and gender-matched non-ALS individuals are included: Participants C (a 65-year-old female) and D (a 79-year-old male), corresponding to Participants A and B, respectively. All participants, B, C, and D, adhere to the same protocol as Participant A when typing the training sentence involving 15 sequences per character. However, during testing, the number of sequences per character varies: 8 for Participant B, 3 for Participant C, and 4 for Participant D. Figure \ref{fig: participants weights} illustrates the estimated contribution weights of the four participants. The result consistently underscores the importance of the parietal and occipital regions in P300 BCI tasks. With the criteria being a posterior selection probability larger than 90\%, channels PO7, PO8, and Oz are consistently selected across all participants, while channel Pz is selected for all except Participant D. Only Participant B selects channel T8. 

\begin{table}[h]
    \centering
    \caption{Demographics and EEG channel selection for the four participants in the multi-participant study. Channels with a posterior inclusion probability of over 90\% are selected. }
    \label{tab: participants demog}
    \begin{tabular}{c|c|c|c|c}\midrule
    Participant&Age&Gender&ALS&Selected Channels\\\midrule
    A&68&F&Yes&PO8, Oz, PO7, Pz\\\midrule
    B&66&M&Yes&PO7, PO8, Oz, Pz, T8\\\midrule
    C&65&F&No&PO8, Oz, PO7, Pz\\\midrule
    D&79&M&No&Oz, PO8, PO7\\\midrule
    \end{tabular}
    
\end{table}

Table \ref{tab: participants performance} shows the BCI performance of different methods for the four participants in both ``Standard Training" and ``Less Training" scenarios. These results further illustrate the effectiveness of GLASS, particularly in scenarios with limited training data. The performance of GLASSel, which uses only channels PO7, PO8, Oz, and Pz, shows a slight decrease compared to the full 16-channel GLASS in the case of Participant B. This could be attributed to the exclusion of channel T8, which is one of Participant B’s selected channels. Despite this, GLASSel still surpasses competing methods in both prediction accuracy and BCI utility. From a practical standpoint, GLASSel presents an optimized set of selected channels, offering potential benefits in reducing electrode costs and simplifying the manufacturing process.

\begin{table}[H]
    \centering
    \caption{BCI performance of multiple participants, quantified in terms of prediction accuracy (\%) and BCI utility (bits/min). In the ``Standard Training" scenario, all of the 15 sequences per character are used for training the model. The ``Less Training" scenario employs only three sequences per character, resulting in an 80\% reduction in training set size.}
    \label{tab: participants performance}
    \resizebox{0.75\textwidth}{!}{
    
    \begin{tabular}{c|c|c|c|c}
    \multicolumn{5}{c}{Standard Training; A: Prediction Accuracy; U: BCI Utility}\\\toprule
Method&Participant A&Participant B&Participant C&Participant D\\\midrule
GLASS & \makecell{A: 93.2\% (4.7\%) \\ U: 21.6 (3.4)} & \makecell{\textbf{77.1\% (10.4\%)} \\ \textbf{7.6 (2.9)}} & \makecell{91.2\% (3.7\%) \\ 26.6 (3.5)} & \makecell{92.4\% (5.9\%) \\ 25.3 (6.0)}\\
\midrule
GLASSel&\makecell{\textbf{94.4\%(3.3\%)} \\ \textbf{22.0 (3.7)}}&\makecell{72.7\% (10.0\%)\\6.7 (3.3)}&\makecell{\textbf{92.9\% (6.3\%)}\\\textbf{28.8 (4.8)}}&\makecell{\textbf{94.0\% (6.2\%)}\\\textbf{27.5 (7.3)}}\\\midrule
swLDA&\makecell{89.9\% (6.8\%)\\18.1 (4.5)}&\makecell{73.2\% (10.4\%)\\6.5 (3.1)}&\makecell{92.2\% (4.6\%)\\28.1 (4.7)}&\makecell{88.8\% (5.5\%)\\22.5 (5.4)}\\\midrule
Logistic&\makecell{85.3\% (7.4\%)\\17.6 (4.9)}&\makecell{73.4\% (7.5\%)\\6.3 (2.1)}&\makecell{83.6\% (9.6\%)\\20.6 (6.6)}&\makecell{84.4\% (6.8\%)\\18.8 (5.4)}\\\midrule
OLS&\makecell{88.1\% (8.9\%)\\15.7 (4.6)}&\makecell{68.0\% (9.0\%)\\4.7 (2.4)}&\makecell{91.7\% (5.3\%)\\28.3 (5.4)}&\makecell{86.4\% (7.2\%)\\19.7 (5.6)}\\\midrule
EEGNet&\makecell{89.7\% (5.8\%)\\14.9 (4.4)}&\makecell{45.1\% (8.5\%)\\0.4 (0.8)}&\makecell{86.2\% (6.0\%)\\20.1 (5.3)}&\makecell{84.8\% (5.9\%)\\15.4 (4.4)}\\\midrule
SVM&\makecell{77.6\% (11.8\%)\\11.3 (4.6)}&\makecell{46.2\% (9.6\%)\\0.6 (1.2)}&\makecell{82.8\% (9.1\%)\\18.0 (3.5)}&\makecell{80.8\% (9.6\%)\\16.9 (6.9)}\\\midrule
SMGP&\makecell{71.1\% (7.9\%)\\6.5 (2.1)}&\makecell{55.3\% (5.5\%)\\1.6 (1.3)}&\makecell{85.1\% (5.2\%)\\19.6 (3.4)}&\makecell{74.2\% (9.8\%)\\13.2 (5.9)}\\\midrule
XGBoost&\makecell{62.7\% (13.5\%)\\4.0 (4.2)}&\makecell{34.1\% (5.8\%)\\0.0 (0.0)}&\makecell{79.9\% (4.8\%)\\17.8 (3.0)}&\makecell{63.5\% (9.6\%)\\7.6 (4.3)}\\\bottomrule
\multicolumn{5}{c}{}\\
\multicolumn{5}{c}{Less Training; A: Prediction Accuracy; U: BCI Utility}\\\toprule
Method&Participant A&Participant B&Participant C&Participant D\\\midrule
GLASS&\makecell{A: 93.1\% (6.3\%)\\U: 21.1 (5.6)}&\makecell{\textbf{67.6\% (9.9\%)}\\\textbf{4.6 (2.7)}}&\makecell{\textbf{92.2\% (6.4\%)}\\\textbf{27.1 (4.6)}}&\makecell{90.1\% (7.6\%)\\20.6 (5.5)}\\\midrule
GLASSel&\makecell{\textbf{93.5\% (3.7\%)}\\\textbf{21.7 (4.1)}}&\makecell{60.9\% (9.3\%)\\3.4 (2.8)}&\makecell{90.1\% (6.4\%)\\24.9 (4.3)}&\makecell{\textbf{90.4\% (6.3\%)}\\\textbf{21.2 (4.0)}}\\\midrule
swLDA&\makecell{80.4\% (17.0\%)\\11.9 (6.2)}&\makecell{17.3\% (8.8\%)\\0.0 (0.0)}&\makecell{86.0\% (4.9\%)\\23.0 (4.2)}&\makecell{76.3\% (8.8\%)\\13.6 (3.9)}\\\midrule
Logistic&\makecell{32.1\% (20.7\%)\\0.0 (0.0)}&\makecell{27.4\% (11.4\%)\\0.0 (0.0)}&\makecell{72.1\% (8.4\%)\\13.2 (5.2)}&\makecell{54.0\% (9.4\%)\\3.1 (5.4)}\\\midrule
OLS&\makecell{58.5\% (18.9\%)\\4.1 (7.5)}&\makecell{19.7\% (5.9\%)\\0.0 (0.0)}&\makecell{62.9\% (9.7\%)\\8.3 (5.6)}&\makecell{53.7\% (8.0\%)\\2.3 (3.1)}\\\midrule
EEGNet&\makecell{23.4\% (11.0\%)\\0.0 (0.0)}&\makecell{22.8\% (6.6\%)\\0.0 (0.0)}&\makecell{62.7\% (6.9\%)\\5.9 (3.2)}&\makecell{34.7\% (6.4\%)\\0.0 (0.0)}\\\midrule
SVM&\makecell{61.2\% (9.6\%)\\3.3 (1.9)}&\makecell{24.8\% (7.4\%)\\0.0 (0.0)}&\makecell{68.1\% (8.1\%)\\10.8 (4.9)}&\makecell{60.3\% (10.4\%)\\5.5 (6.0)}\\\midrule
SMGP&\makecell{68.8\% (9.2\%)\\6.3 (2.5)}&\makecell{25.4\% (9.2\%)\\0.0 (0.0)}&\makecell{66.5\% (10.8\%)\\8.7 (5.1)}&\makecell{68.5\% (11.5\%)\\9.4 (5.8)}\\\midrule
XGBoost&\makecell{50.1\% (15.0\%)\\1.8 (2.7)}&\makecell{8.9\% (5.9\%)\\0.0 (0.0)}&\makecell{57.4\% (12.4\%)\\4.5 (7.1)}&\makecell{32.3\% (10.3\%)\\0.0 (0.0)}\\\bottomrule
    \end{tabular}
    }
    
\end{table}

\section{Simulations}\label{sec: simulation}

\subsection{Simulation I}

To best mimic the real data, we apply a label-switching method to the real data collected from Participant A. This simulation serves two purposes: (i) to test the estimation precision of parameters in GLASS and (ii) to compare the prediction accuracy of GLASS against other methods. In addition to the ``Standard Training" and the ``Less Training" scenarios, we examine two misspecified scenarios: ``Attention Drift" and ``Noisy EEG". To maintain a signal-to-noise ratio similar to that in the real data, we base the true model on the GLASS model fitted to Participant A's real data. We designate channels Pz, PO7, PO8, and Oz as signal channels ($\delta_e=1$) and the remaining channels as noise channels ($\delta_e=0$). Contribution weights of Pz, PO7, PO8, and Oz are set to be their $L_2$-normalized fitted posterior medians. The true latent channel time-varying effects are set to be their fitted posterior medians.

The simulation procedure is outlined as follows: First, the target row/column number $z_\kappa$ is generated based on \eqref{eq: linear}. If the generated $z_\kappa$ differs from the one in the real data, we switch the two row/column numbers. For instance, if row 1 is the target in the real data, but the simulated target row number is 3, we relabel the original row 1 as row 3 and the original row 3 as row 1. This ensures that all half-sequences for the same character have the same simulated target row/column number. Additionally, we randomly permute the row/column numbers of non-target stimuli in each half-sequence to introduce more randomness to the simulation.

The ``Standard Training" and ``Less Training" scenarios are defined as the same in the application study. In the ``Standard Training" scenario, we fit each model with the entire training dataset, while in the ``Less Training" scenario, we only take the first three of the fifteen sequences per character for model training, reducing the training size by 80\%. We consider two additional mis-specified scenarios for a comprehensive simulation study. In the ``Attention Drift" scenario, there is a 10\% chance of randomly swapping a target stimulus with a non-target one. The ``Noisy EEG" setting introduces extra noise to each EEG channel in both training and testing datasets. The noise follows an autoregressive distribution with an autoregressive coefficient of 0.5 and a variance of 1. All scenarios are repeated 50 times for robustness.

The model fitting configurations are identical to these in Section \ref{sec: application}. To evaluate the estimation accuracy of latent channel time-varying effects, we define the relative mean square error ($RMSE$) to be $RMSE=||\Tilde{\bfbeta}-\Tilde{\bfbeta}_{esti}||^2/||\Tilde{\bfbeta}||^2$, where $\Tilde{\bfbeta}$ and $\Tilde{\bfbeta}_{esti}$ are the true and estimated column vector of latent channel time-varying effects, respectively. An $RMSE$ value close to zero implies an accurate estimation of latent channel time-varying effects. We compute the average posterior selection probabilities of signaled channels and noise channels, denoted as $\hat{\delta}_{signal}$ and $\hat{\delta}_{noise}$, respectively. Higher $\hat{\delta}_{signal}$ and lower $\hat{\delta}_{noise}$ imply correct channel selection. We also evaluated the estimation accuracy of the contribution weights of channels. The true vector of contribution weights, denoted as $\bfalpha$, is a 16-dimensional unit vector, but the posterior median of $\bfalpha$ may not be a unit vector. Therefore, we estimate $\bfalpha$ using the $L_2$-normalized posterior median vector, denoted as $\hat{\bfalpha}$. The estimation accuracy of $\bfalpha$ is measured by the angle between $\bfalpha$ and $\hat{\bfalpha}$. 

\begin{table}[h]
\centering
\caption{Key metrics for assessing the parameter estimation accuracy of GLASS across different scenarios. The relative mean square error (RMSE) evaluates the accuracy of estimating latent channel time-varying effects, with lower values indicating higher accuracy. $\hat{\delta}_{signal}$ and $\hat{\delta}_{noise}$ represent the average posterior selection probabilities of signal and noise channels, respectively. The error angle, measured between the true and estimated unit vectors of contribution weights, quantifies the accuracy in estimating channel contributions. The ``Standard Training" scenario utilizes the correctly specified simulation dataset for model fitting, while the ``Less Training" scenario reduces the training data size by 80\%. Additionally, ``Attention Drift" and ``Noisy EEG" represent two misspecified scenarios.}
\label{tb::sim1 esti}
\begin{tabular}{c|c|c|c|c}
\midrule
& Standard Training & Less Training & Attention Drift & Noisy EEG \\
\midrule
RMSE &7.3\% (1.4\%) & 38.0\% (13.8\%) & 13.0\% (3.8\%) & 15.4\% (2.8\%) \\\midrule
$\hat{\delta}_{signal}$& 93.0\% (2.4\%) & 86.4\% (8.1\%) & 92.1\% (2.5\%) & 91.6\% (2.6\%) \\\midrule
$\hat{\delta}_{noise}$& 37.8\% (5.4\%) & 41.5\% (5.4\%) & 39.0\% (5.8\%) & 39.7\% (5.6\%) \\\midrule
Error Angle& 3.1° (0.7°) & 8.4° (2.6°) & 4.3° (2.3°) & 5.2° (1.6°) \\\midrule
\end{tabular}
\end{table}

\begin{figure}[h]
    \centering
    \includegraphics[width=1\linewidth]{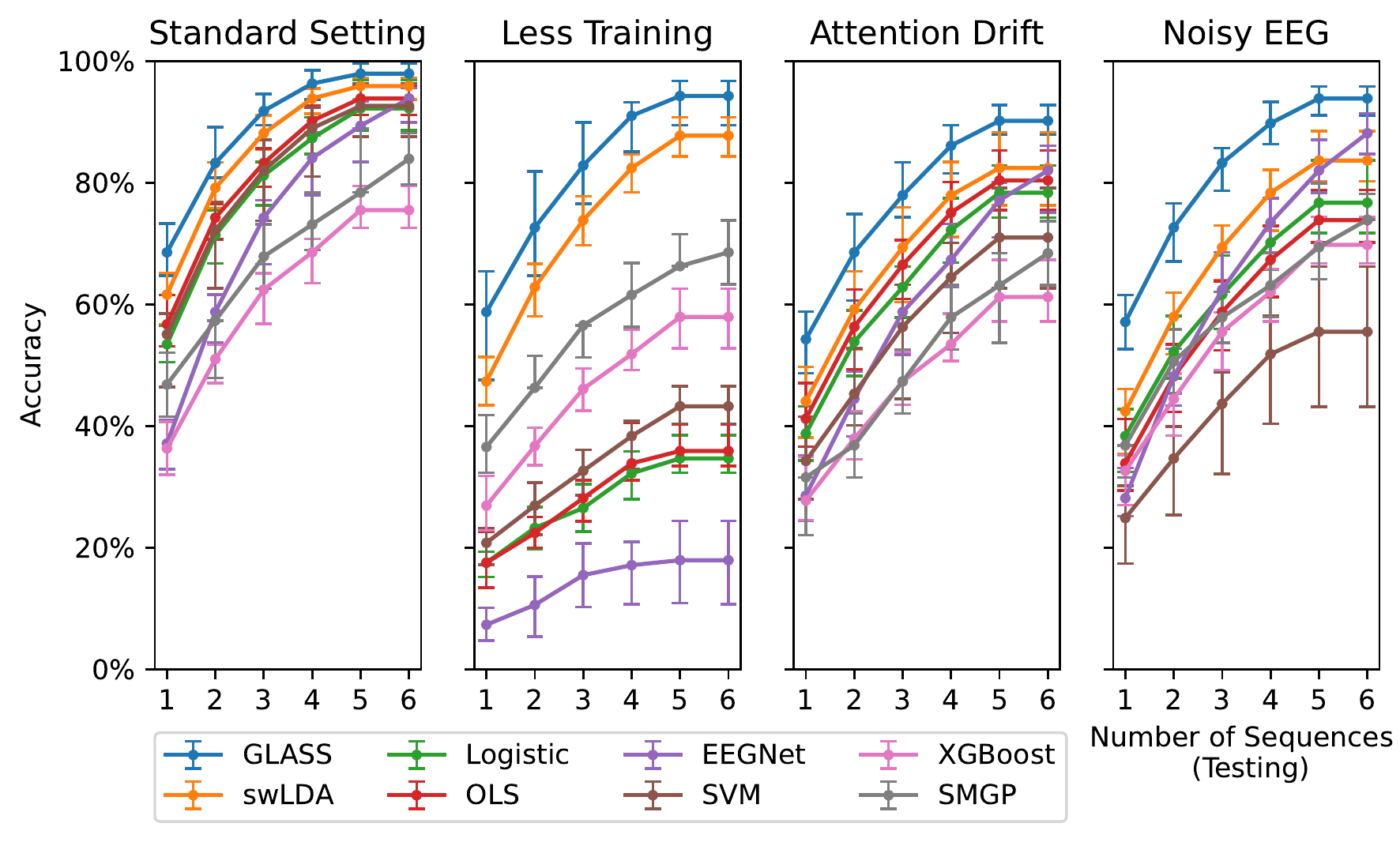}
    \caption{Prediction accuracy of GLASS compared to other competing methods across various training scenarios. In the ``Standard Training" scenario, the correctly specified simulation dataset is used for model fitting. The ``Less Training" scenario uses only 80\% of the training data for model fitting. The figure also includes results from two misspecified scenarios: ``Attention Drift" and ``Noisy EEG"}
    \label{fig:sim1 accu}
\end{figure}

Table \ref{tb::sim1 esti} presents results for the parameter estimation accuracy of GLASS. Under conditions simulating a real-data-like signal-to-noise ratio, GLASS demonstrates correct identification of signal channels and accurately estimates latent channel time-varying effects and contribution weights. GLASS maintains a moderate reduction in performance under misspecified conditions, indicating its robustness. Figure \ref{fig:sim1 accu} shows the prediction accuracy of GLASS and competing methods. GLASS outperforms competing methods, exhibiting higher prediction accuracy and lower standard deviation. The advantage of GLASS is particularly evident in the scenario with limited training data, highlighting the potential of GLASS in reducing the calibration effort of BCIs. The advantage of GLASS is particularly pronounced in the scenario with limited training data, underscoring its potential to reduce the calibration effort in BCI applications.

\subsection{Simulation II}

This simulation is a replication of the simulation study from \cite{ma2022bayesian}. We compare the prediction accuracy of GLASS with other competing methods in this simulation. This simulation represents a misspecified scenario for GLASS, as the underlying true model is the Bayesian generative model SMGP, differing from the model assumptions of GLASS.

The simulation employs three EEG channels operating at a frequency of 32 Hz. The simulated stimulus presentation adopts the same scheme as in our study, which involves 12 stimuli per sequence corresponding to each row and column on the $6\times6$ keyboard. Their design includes 19 characters and five sequences per character for both training and testing datasets. The EEG signal simulation process involves first generating background EEG noise, characterized by an auto-regressive temporal structure and a compound symmetry spatial dependency. The variance of the background noise, denoted as $\sigma_x^2$, is set at two different levels to simulate varying noise conditions: 20 for a moderate noise level and 40 for a high noise level. Subsequently, ERP responses are added to these background EEG noises following each stimulus. Different ERP functions are used for target and non-target stimuli, as detailed in the appendix files of \cite{ma2022bayesian}. 

\begin{table}
\centering
\caption{Averaged character-level accuracy and standard errors over 100 simulations. The variance of the background EEG noise, $\sigma_x^2$, is set at two levels: 20 for a moderate noise level and 40 for a high noise level. The results for the first seven methods are initially reported in \cite{ma2022bayesian}, while the results for the last two methods are contributions exclusive to our study.}
\label{tab::sim smgp}
\begin{tabular}{l|c|c|c|c|c|c}
\midrule
\multirow{2}{*}{\diagbox[width=2cm]{Mthd}{$n_{seq}$}} & 3 & 4 & 5 & 3 & 4 & 5 \\
\cline{2-7}
&\multicolumn{3}{c|}{$\sigma_x^2=20$}& \multicolumn{3}{c}{$\sigma_x^2=40$}\\
\midrule
SMGP & 0.91 (0.07) & 0.96 (0.04) & 0.99 (0.03) & 0.67 (0.11) & 0.79 (0.09) & 0.87 (0.08) \\
NN & 0.76 (0.10) & 0.87 (0.08) & 0.92 (0.07) & 0.55 (0.12) & 0.66 (0.11) & 0.75 (0.10) \\
SVM & 0.81 (0.09) & 0.89 (0.07) & 0.94 (0.06) & 0.54 (0.11) & 0.64 (0.12) & 0.73 (0.10) \\
Logistic & 0.76 (0.08) & 0.87 (0.07) & 0.91 (0.06) & 0.53 (0.11) & 0.62 (0.11) & 0.71 (0.11) \\
RF & 0.76 (0.10) & 0.86 (0.08) & 0.92 (0.06) & 0.50 (0.12) & 0.62 (0.12) & 0.69 (0.11) \\
swLDA & 0.85 (0.08) & 0.93 (0.06) & 0.97 (0.04) & 0.59 (0.11) & 0.71 (0.11) & 0.80 (0.10) \\
XGBoost & 0.67 (0.11) & 0.77 (0.09) & 0.85 (0.08) & 0.45 (0.12) & 0.56 (0.11) & 0.65 (0.11) \\
GLASS&0.89 (0.03)&0.96 (0.04)&0.97 (0.04)&0.64 (0.11)&0.78 (0.11)&0.83 (0.10)\\
OLS&0.70 (0.11)&0.81 (0.08)&0.89 (0.07)&0.52 (0.13)&0.67 (0.11)&0.77 (0.12)\\
\midrule
\end{tabular}
\end{table}

Table \ref{tab::sim smgp} reports the averaged character-level accuracy and standard errors across 100 simulated datasets. In this generative simulation, where SMGP serves as the true model, GLASS not only achieves accuracy comparable to SMGP but also outperforms other competing methods in terms of prediction accuracy and stability. This simulation adopts a relatively low sampling rate of 32 Hz, which implies a low spatial resolution. The results indicate that GLASS is also suitable for applications that involve low-frequency EEG signals, demonstrating its versatility and effectiveness in various EEG signal scenarios.

\section{Conclusion}
\label{sec: conc}

In this study, we introduced a novel Bayesian multinomial logistic regression model tailored specifically for P300 BCI applications. Our multinomial regression framework is fundamentally different from traditional binary classification methods. To the best of our knowledge, this is the first work to explicitly address the dataset imbalance issue inherent to the P300 BCI design. The introduction of a latent channel successfully addresses spatial correlations in EEG data while summarizing information from the original channels. The soft-thresholded GP prior plays a critical role in feature selection and noise reduction, significantly enhancing the SNR of EEG data. The unique strengths of GLASS, our proposed model, are highlighted through comprehensive simulation studies and real-world applications. Our work not only paves the way for more advanced, data-driven approaches in brain-computer interfacing but also opens up new possibilities for enhancing BCI technologies and gaining deeper insights into EEG data analysis.









\bibliographystyle{agsm}

\bibliography{main}
\end{document}